\documentclass[sigconf,nonacm]{acmart}

\setcopyright{none}

\usepackage{cleveref}
\usepackage{subcaption}
\usepackage{amsmath}
\usepackage{amsfonts}
\usepackage{dsfont}
\usepackage{multirow}

\def\crash{\text{Crash}}
\def\crashprimary{\text{Crash}^{(0.20)}}
\def\retex{R^{ex}}
\def\ESG{\text{ESG}}
\def\stress{\text{Stress}}

\begin{document}

\title[ESG as Priced Crash Insurance]{ESG as Priced Crash Insurance: State-Dependent Tail Risk and Deconfounding Evidence}

\author{Jiayu Yi}
\authornote{Corresponding Author}
\orcid{0000-0003-1619-1035}
\affiliation{%
  \institution{Nanyang Technological University}
  \city{Singapore}
  \country{Singapore}
}
\email{sophiayi97@gmail.com}

\author{Minxuan Hu}
\orcid{0009-0007-1733-0099}
\affiliation{%
  \institution{Cornell University}
  \city{Ithaca}
  \state{NY}
  \country{USA}
}
\email{mh2229@cornell.edu}

\author{Wenxi Sun}
\orcid{0009-0007-2585-6015}
\affiliation{%
  \institution{Johns Hopkins University}
  \city{Baltimore}
  \state{MD}
  \country{USA}
}
\email{wsun41@alumni.jh.edu}

\author{Ziheng Chen}
\orcid{0000-0001-9671-3977}
\affiliation{%
  \institution{University of Texas at Austin}
  \city{Austin}
  \state{TX}
  \country{USA}
}
\email{stokes615@utexas.edu}

\renewcommand{\shortauthors}{Yi et al.}

\begin{abstract}

This research establishes ESG as a state dependent insurance mechanism against equity crashes by addressing the decoupling of unconditional alpha from tail risk resilience. By validating market stress regimes as distinct economic states through a drawdown based truncation rule, the study demonstrates that high ESG ratings materially reduce the incidence of discrete crash events during systemic drawdowns. To address the selection bias and high dimensional confounding inherent in traditional linear frameworks, we implement Double Machine Learning as a structural deconfounding layer. Unlike simple predictive modeling, the Double Machine Learning framework utilizes machine learning to handle complex nuisance parameters, allowing us to isolate the asymmetric treatment effects of ESG across different market states. Distributional analysis reveals the underlying mechanism as ESG specifically attenuates the severity of realized tail losses at the most adverse quantiles instead of shifting the entire return distribution. Confirmed by structural estimates, this protection functions as priced insurance that incurs performance drags during stable periods while providing critical resilience when tail risks are most acute.

\end{abstract}



\maketitle

\section{Introduction}

The pervasive integration of Environmental, Social, and Governance (ESG) criteria in recent years has fundamentally restructured the operational paradigms of institutional risk management. While significant progress has been made in identifying average return premiums, the application of such ratings to the management of extreme market drawdowns remains a critical frontier. At the heart of this debate is whether ESG acts as a vehicle for positive screening to capture unconditional Alpha or as a state dependent insurance mechanism designed for tail risk hedging. A persistent challenge in current sustainable finance is the methodological divergence between pricing model calibration often focused on unconditional alpha and actual downside protection performance. 
Traditional frameworks often rely on static specifications or latent-state models that can blur abrupt nonlinear shifts in market conditions.
Such linear assumptions often mask the true value of ESG as its protective effects are typically dormant during stable periods and only activate under systemic pressure, creating a fundamental state dependent necessity for any empirical investigation.

To address these challenges, this research implements a multi stage empirical framework designed to isolate the state dependent properties of ESG. The study first establishes an objective market stress indicator by applying a drawdown based truncation rule to identify systemic volatility regimes. We then employ regime specific logit models to analyze the incidence of discrete crash events across different market states while utilizing conditional quantile regressions to map the association between ESG ratings and extreme left tail return distributions. Crucially, the methodology incorporates Double Machine Learning as a structural deconfounding layer to handle high dimensional firm characteristics and potential selection bias. This approach utilizes flexible machine learning estimators to partial out nuisance parameters and identify the treatment effect of ESG scores on return resilience with higher causal rigor than traditional linear specifications.

Our analysis reveals that the protective effect of ESG is strictly state contingent and concentrated in the ultra left tail of the realized return distribution. Instead of acting as a broad performance driver, high ESG ratings specifically reduce the odds of crossing discrete crash thresholds during systemic drawdowns. Quantile evidence demonstrates that the ESG association is statistically concentrated at crash like quantiles during stress months to explain the lower frequency of discrete crash events.  Structural estimates from the Double Machine Learning deconfounding layer further confirm that these patterns remain asymmetric across states and act as a priced insurance mechanism instead of a driver of excess returns. Within this framework, the Social pillar emerges as the primary resilience driver while the Environmental pillar represents the cost of insurance through performance drags during stable periods.


This paper delivers three primary research contributions. (1) We move toward a drawdown-based market-stress indicator that provides a transparent and empirically validated state object for the analysis of ESG under tail risk. This state-contingent framework underpins all subsequent analyses and enables us to study ESG protection under explicitly adverse market conditions. (2) We introduce state dependent crash models to demonstrate that high ESG ratings are associated with a lower incidence of discrete crash events during market turbulence. By addressing the selection bias and high dimensional confounding inherent in linear frameworks, we implement Double Machine Learning as a structural deconfounding layer to isolate the asymmetric treatment effects of ESG across different market states. (3) We deepen the analysis into tail severity to reveal that ESG protective effects are heavily concentrated in the extreme lower tails of the return distribution during stress months. This structural evidence confirms that ESG functions as priced insurance that handles complex nuisance parameters to provide critical resilience precisely when systemic fragility is high.

The remainder of this paper is structured as follows: Section 2 provides a comprehensive review of the relevant literature; Section 3 describes the firm-month panel data and variables; Section 4 details the empirical methodology; Section 5 presents the empirical analysis, covering market stress indicator and validation, state-dependent crash-event incidence, excess-return tail severity in stress regimes, and deconfounded ESG effects across
stress states; Section 6 conducts robustness checks of crash threshold sensitivity and pillar score effects; and Section 7 concludes.

\section{Literature Review}

One important debate within the literature on ESG relates to whether or not the high ESG ratings imply that investors will receive unconditional return premia or simply provide state-contingent protection.
For example, the performance of publicly traded U.S. firms is studied during the financial crisis of 2008-2009 \cite{lins2017social}  and it turns out that companies with high CSR generated stock returns that were four to seven percentage points higher than those of firms with low CSR.
This outperformance supports the "Stakeholder Theory" \cite{freeman2010strategic} where ESG engagement functions as an investment in Social Capital.
Such accumulated trust remains dormant during stable periods but activates as a critical survival mechanism during systemic ruptures by providing a liquidity buffer that prevents fire sale dynamics and mitigates equity crash risk.
Consequently, the value of ESG is derived from this stakeholder trust being uniquely priced during market distress rather than acting as a general driver of returns.

Complementing the trust-based view, the Governance and Environmental pillars address the agency cost perspective of market crashes \cite{jin2006r2}.
Financial theory suggests that price crashes often result from managers hoarding negative information to protect their private benefits \cite{hutton2009opaque}.
Robust ESG frameworks impose stricter monitoring and transparency that reduce the managerial incentive to conceal bad news and ensure a more continuous flow of information to the market, which prevents the discrete catastrophic price adjustments that occur when accumulated negative news reaches a breaking point.
Demers et al. \cite{demers2021esg} issue an important correction to this argument: after carefully controlling for industry affiliation, market-based risk measures, and investments in intangible assets, ESG does not provide significant predictive value for returns during the COVID-19 pandemic.
The study demonstrates that the apparent ``protection'' from ESG measures probably reflects correlations between firms' characteristics rather than actual causal relationships.
This criticism aligns with recent advances in machine learning for asset pricing where high-dimensional characteristics often exhibit complex nonlinear interactions \cite{gu2020empirical}.
Specifically, a major pain point in empirical finance is the high degree of multicollinearity between ESG scores and fundamental features such as size, momentum, and volatility, which often renders traditional linear models incapable of identifying a pure ESG effect.
Therefore, we intend to employ a high-dimensional deconfounding method to test this proposition using the data available to us as of October, 2025.

To tackle this identification issue, we use Double Machine Learning (DML), a framework developed by Chernozhukov et al. \cite{chernozhukov2018double}.
DML effectively neutralizes the regularization bias and overfitting issues inherent in flexible machine learning estimators through the incorporation of Neyman-orthogonal scores and cross-fitting.
Thus, we are able to obtain root-N consistency and asymptotic normality in our estimates on the effect of ESG, providing a reliable statistical inference.
By utilizing Double Machine Learning as a structural deconfounding layer, we isolate the pure ESG signal from high-dimensional noise and firm-specific confounders to determine whether the insurance effect derived from reduced agency costs persists once traditional financial resilience is accounted for.

Finally, we use Koenker and Bassett's conditional quantile regression \cite{koenker1978regression} to identify the position of ESG protection within its distribution of returns.
Conditional quantiles can characterize all of a conditional distribution's quantiles without any parametric restrictions or assumptions which makes them well-suited to detecting left-tail protection effects that would not be observed using mean regression.
This is economically essential because if ESG functions as priced insurance its impact should attenuate the severity of realized losses at the most adverse quantiles during stress regimes rather than shifting the entire return distribution.

In several disciplines, recovering significant signals that are hidden due to the presence of high-dimensional noise is a constant theme. For example, in e-commerce security, many bots can only be identified by their anomalous relational patterns contained within session–URL graph structures even though their individual behavioural features are considered normal \cite{zhao2026non}.
Similarly, structures of risk caused by logistics congestion and other variables can only emerge once heterogeneous sensor noise has undergone spatiotemporal filtering \cite{xue2026resilient}.
The DML framework presented here mimics this same principle of determining a signal from noise in regard to the protection against ESG tail risk; in doing so, the DML framework conditions upon different market stress regimes while controlling for firm-specific confounding factors, allowing for the identification of a protective relationship between ESG and tail risk that would otherwise not be visible in unconditional mean regression analyses.

\section{Data and Variables}

\subsection{Sample Selection and Universe Construction}

Our primary sample consists of a firm-month panel of U.S. equities constituents of the S\&P 500 index spanning the period from 2015 to 2025.
To mitigate survivorship bias, we construct a time-varying contemporaneous universe by tracking index additions and deletions on each trading day.
The raw data includes daily closing prices adjusted for corporate actions (e.g., splits and dividends), from which we derive monthly total returns ($r_{i,t}$) as the primary dependent variable \cite{compustat}.

\subsection{ESG Ratings and Pillar Scores}

The key explanatory variable is the Environmental, Social, and Governance (ESG) rating provided by MSCI \cite{esg}.
These ratings quantify a firm's resilience to long-term, financially material ESG risks and opportunities relative to industry peers.
We utilize the numerical Industry Adjusted Score ($D_{i,t} := ESG_{i,t-1}$) to ensure granular precision in our econometric modeling.
For the pillar-level heterogeneity analysis discussed in \cref{subsec:pillar-dml}, we further decompose the aggregate score into its individual Environmental (E), Social (S), and Governance (G) components.

\subsection{Firm-Level Control Variables}

To isolate the causal impact of ESG and address potential confounding with established financial characteristics, we incorporate a high-dimensional vector of lagged firm controls ($X_{i,t-1}$) \cite{compustat}.
We select variables with a missing-data rate of less than 20\% and apply standard normality corrections and standardization to ensure data quality.
The final set of controls includes size, leverage, profitability, investment, tangibility, and sector fixed effects. 
A summary table is provided at \Cref{tab:variable_definitions}.

\begin{table*}[htbp]
\centering
\caption{Variable Definitions and Specification Details.}
\label{tab:variable_definitions}
\begin{tabular*}{\textwidth}{@{\extracolsep{\fill}} lllc @{}}
\toprule
\textbf{Category} & \textbf{Symbol} & \textbf{Variable Construction / Definition} & \textbf{Source} \\
\midrule
Dependent Variable & $R_{i,t}^{ex}$ & $\text{Excess Return} = r_{i,t} - r_{m,t}$ & Compustat \cite{compustat}\\
Main Treatment & $\text{ESG}_{i,t-1}$ & Industry Adjusted Aggregated ESG Score & MSCI \cite{esg}  \\
Market Regime & $\text{Stress}_t$ & $\mathbb{I}\{r_{m,t} \leq q_{0.15}(r_m)\}$ (Bottom 15\% of market returns) & Calculated  \\
Crash Event & $\text{Crash}_{i,t}^{(0.20)}$ & $\mathbb{I}\{r_{i,t} < -0.20\}$ (Discrete 20\% drawdown) & Calculated  \\
\midrule
\textbf{Firm Controls} & $X_{i,t-1}$& & Compustat \cite{compustat} \\
Size & $\text{log}(\text{at})$ & Logarithm of total assets &  \\
Leverage & $\text{lev}$ & Total long-term debt / Total assets ($\text{dltt}/\text{at}$) &  \\
Profitability & $\text{prof}$ & Income before extraordinary items / Total assets ($\text{ib}/\text{at}$) &  \\
Investment & $\text{inv}$ & Capital expenditures / Total assets ($\text{capx}/\text{at}$) &  \\
Tangibility & $\text{tang}$ & Net property, plant, and equipment / Total assets ($\text{ppent}/\text{at}$) &  \\
\bottomrule
\end{tabular*}
\end{table*}

\section{Methodology}
\subsection{Market Stress Indicator}
\label{sec:method:stress}

Our hypotheses are state-dependent: ESG should matter most when market conditions deteriorate. 
To classify market conditions, we construct a monthly market return from the firm-month panel
\begin{align}
r_{m,t}&:=\sum_{i\in\mathcal{I}_t} \widetilde{w}_{i,t-1}\, r_{i,t}
\end{align}
under the weighting scheme
\begin{align}
\widetilde{w}_{i,t-1}:=\frac{w_{i,t-1}}{\sum_{j\in\mathcal{I}_t} w_{j,t-1}}, \quad
 w_{i,t-1} := \frac{V_{i,t-1}^{1/3}}{\sigma_{i,t-1}}
\end{align}
where $V$ and $\sigma$ stand for trading volume (in dollar units) and volatility and $\mathcal{I}_t$ is the set of firms with non-missing returns and weights in month $t$.
We then define
\[
Stress_t=\mathds{1}\!\left\{r_{m,t}\le q_{0.15}(r_m)\right\},
\]
so that stress months are the bottom 15\% of the monthly market-return distribution and all remaining months are non-stress.
This produces a single monthly regime indicator, $Stress_t$, shared across the crash logits, tail regression, and deconfounding analysis.





\subsection{Crash-Event Modeling by Market Regime}
\label{sec:method:crashlogit}

To study whether ESG is associated with the incidence of discrete crash events, we estimate crash logits separately for stress and non-stress months.

For $c\in\{0.15,0.20,0.25\}$, define
$\crash^{(c)}_{i,t}=\mathds{1}\{r_{i,t}<-c\}$.
Our primary outcome is $\crash_{i,t}\equiv\crash^{(0.20)}_{i,t}$, with $c\in\{0.15,0.25\}$ used for robustness.

For each state $s\in\{0,1\}$ implied by $Stress_t$, we estimate two nested specifications. The first is a parsimonious sector fixed-effects logit:
\begin{equation}
\begin{split}
\Pr\! & \left(\crash_{i,t}=1 \mid \ESG_{i,t-1},\text{sector}(i),\stress_t=s\right) \\
& =
\text{logit}^{-1}\!\Big(
\alpha_s
+\lambda_{s,\text{sector}(i)}
+\beta_s\,\ESG_{i,t-1}
\Big),
\label{eq:logit_regime_a}
\end{split}
\end{equation}
and the second adds lagged firm controls:
\begin{equation}
\begin{split}
\Pr\! & \left(\crash_{i,t}=1 \mid \ESG_{i,t-1},X_{i,t-1},\text{sector}(i),\stress_t=s\right) \\
& =
\text{logit}^{-1}\!\Big(
\alpha_s
+\lambda_{s,\text{sector}(i)}
+\beta_s\,\ESG_{i,t-1} + \theta_s'X_{i,t-1}
\Big).
\label{eq:logit_regime_b}
\end{split}
\end{equation}

Spec.~A~\eqref{eq:logit_regime_a} provides a transparent state-specific baseline, while Spec.~B~\eqref{eq:logit_regime_b} tests whether the same relation survives conditioning on standard firm characteristics. In both cases, $\beta_s$ is the ESG slope within state $s$. Standard errors are clustered by month to account for within-month dependence from common shocks \cite{cameron2015practitioner}.

\subsection{Tail Severity via Conditional Quantile Regression}
\label{sec:method:quantile}

We then examine the same question on the continuous return scale by studying tail severity in firm-level excess returns. Using the same monthly market return $r_{m,t}$, we define excess return as $\retex_{i,t}=r_{i,t}-r_{m,t}$.

Conditional quantile regressions \cite{koenker1978regression} are used to test whether ESG is associated with the left tail of $\retex_{i,t}$, and whether that association differs in stress months. Here $Q_{\tau}(\retex_{i,t}\mid\cdot)$ denotes the conditional $\tau$-th quantile of excess returns, so smaller values of $\tau$ correspond to more extreme left-tail outcomes.

For $\tau\in\{0.01,0.02,0.05,0.10,0.20\}$, we estimate
\begin{equation}
\begin{split}
Q_{\tau}\! & \left(\retex_{i,t}\mid \ESG_{i,t-1},\stress_t,X_{i,t-1},\text{sector}(i)\right) \\
& \sim
\alpha_{\tau}
+\lambda_{\tau,\text{sector}(i)}
+\beta_{\tau}\ESG_{i,t-1}
+\phi_{\tau}\stress_t \\
& \quad + \delta_{\tau}\!\left(\ESG_{i,t-1}\times\stress_t\right)
+\theta_{\tau}'X_{i,t-1},
\label{eq:quantile_main}
\end{split}
\end{equation}
where $\ESG_{i,t-1}$ is the lagged ESG score, $X_{i,t-1}$ collects lagged firm characteristics, and $\lambda_{\tau,\text{sector}(i)}$ are sector fixed effects. The coefficient $\beta_{\tau}$ captures the ESG slope in non-stress months, while $\beta_{\tau}+\delta_{\tau}$ gives the ESG slope in stress months.

We report percentile-based 95\% confidence intervals from a month-block bootstrap that resamples entire months to respect within-month dependence, following the block-bootstrap logic for dependent data \citep{kunsch1989jackknife}. Because stress months are relatively scarce, bootstrap draws resample stress and non-stress months separately and preserve the number of months in each regime.

\subsection{Deconfounding via Double Machine Learning}
\label{subsec:dml-setup}

We define the DML setup using a partialling-out approach with a continuous treatment \cite{chernozhukov2018double}. Our target outcome variable, $Y_{i,t}$, takes two primary forms: the excess return $\retex_{i,t}$ and the crash indicator $\crashprimary_{i,t}$.
Our treatment variable is defined as the lagged ESG score, $D_{i,t} := \ESG_{i,t-1}$.
The high-dimensional control vector, $W_{i,t-1}$, encompasses our standard firm controls ($X_{i,t-1}$), sector assignments, and optionally their flexible non-linear interactions.
To effectively isolate the causal parameter while maintaining valid inference, the DML procedure relies on sample splitting and cross-fitting.
In the first stage, we fit two conditional expectations (nuisance parameters) using flexible machine learning algorithms, such as Lasso, Random Forest (RF), or Gradient Boosting Machines (GBM).
We model the conditional expectation of the outcome and the ESG treatment given the same set of controls:
\begin{align}
\widehat{m}(W) := \mathbb{E}[Y|W], \quad \widehat{g}(W) := \mathbb{E}[D|W].
\label{eq:45-dml-expectation}
\end{align}
Following the estimation of these nuisance functions, we residualize both our outcome and treatment variables to partial out the confounding effects.
The orthogonalized components are calculated as 
$
\widetilde{Y} := Y - \widehat{m}(W), \widetilde{D} := D - \widehat{g}(W).
$
In the final step, we estimate our target coefficient, $\beta$, by regressing the residualized outcome on the residualized treatment:
\begin{align}
\widetilde{Y} \sim \beta \widetilde{D} + \epsilon,\quad \mathbb{E}[\epsilon|\widetilde{D}]=0.
\label{eq:45-dml-final-regression}
\end{align}
By demonstrating that the ESG coefficient $\beta$ remains significant and economically meaningful through this rigorous procedure, we effectively neutralize the objection that downside protection is merely an artifact of linear proxies for size or quality. 



\section{Empirical Results and Discussion}
\subsection{Market Stress Indicator and Validation}
\label{sec:results:stress}

Applying the stress rule defined in Section~\ref{sec:method:stress} yields a cutoff of $-2.82\%$ and classifies 22 of 143 months (15.4\%) as stress months.

\begin{figure*}[!h]
\centering
\begin{subfigure}[h]{0.49\linewidth}
    \centering
    \includegraphics[width=\linewidth]{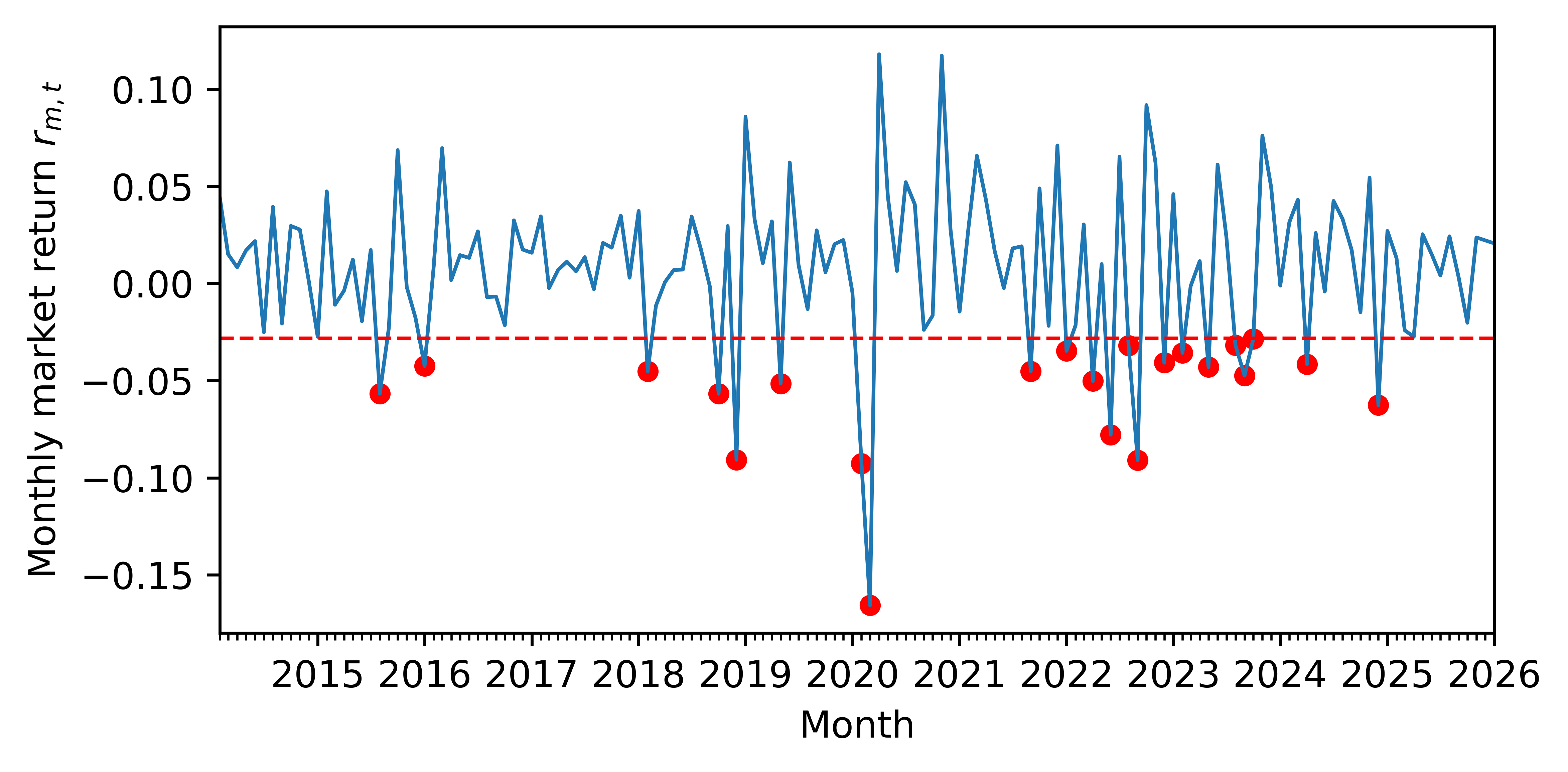}
    \caption{Time Series of Monthly Market Returns.}
    \label{fig:stress_ts}
\end{subfigure}\hfill
\begin{subfigure}[h]{0.50\linewidth}
    \centering
    \includegraphics[width=\linewidth]{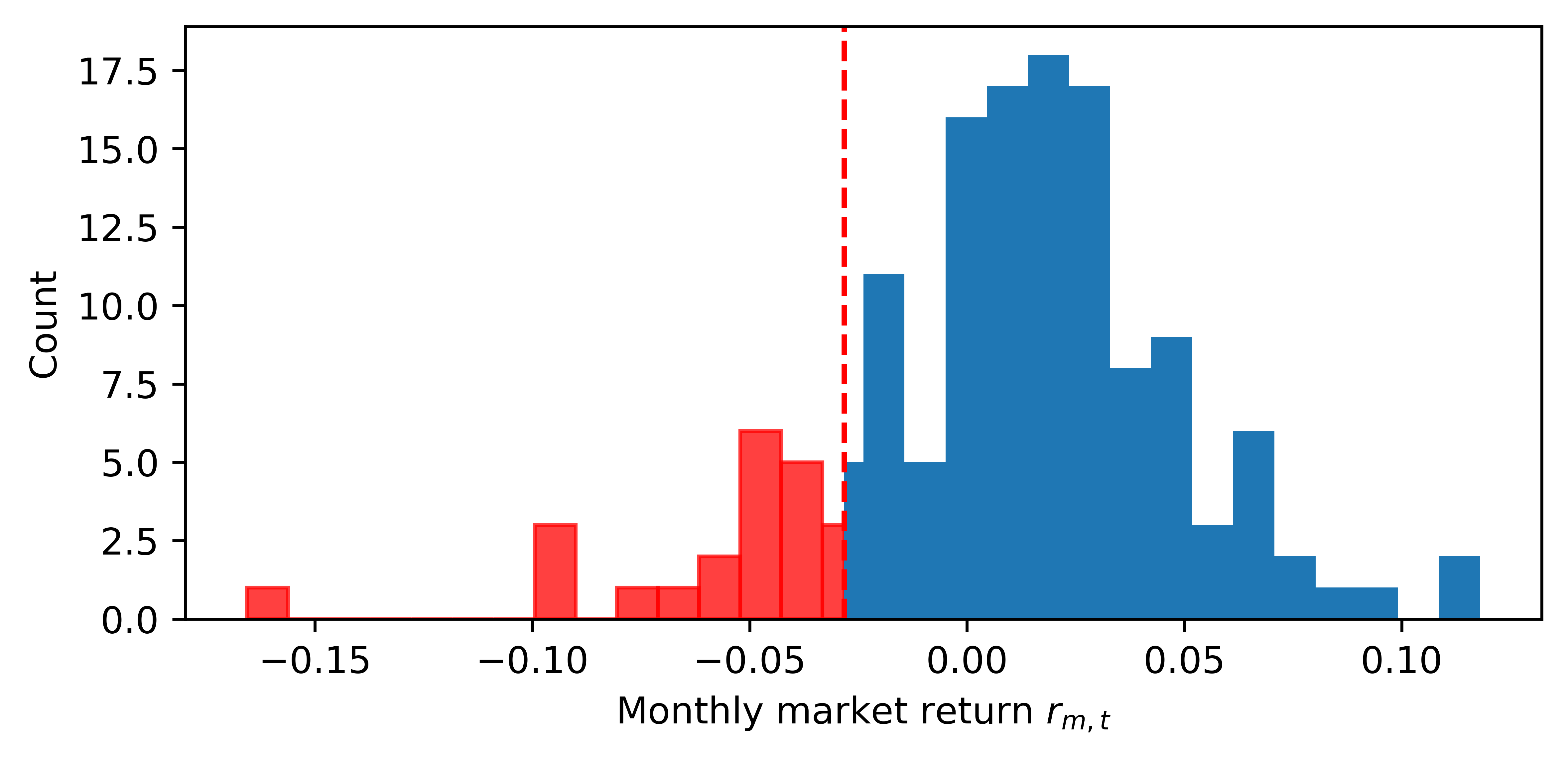}
    \caption{Distribution of Monthly Market Returns.}
    \label{fig:stress_hist}
\end{subfigure}
\caption{Market Stress Definition.}
\vspace{10pt}
\raggedright
\small
\footnotesize
\textit{Notes:} The dashed lines mark the 15th-percentile cutoff ($-2.82\%$). Panel (a) shows monthly market returns with stress months highlighted; panel (b) shows the corresponding distribution.
\label{fig:stress_validation}
\end{figure*}

Figure~\ref{fig:stress_validation} validates that this rule captures broad market drawdowns rather than idiosyncratic noise. The time-series panel shows that stress months coincide with widely recognized market selloffs, including the COVID shock (March 2020: $-16.6\%$; February 2020: $-9.3\%$) and several sharp drawdowns during 2022. The histogram panel confirms that, by construction, the cutoff isolates the left tail of the market-return distribution.

\subsection{State-Dependent Crash-Event Incidence}
\label{sec:results:crashlogit}

Table~\ref{tab:logit_crash20_interaction} presents the headline crash-risk result on the extensive margin. For the primary outcome $\crashprimary_{i,t}$, the ESG slope is negative and statistically significant in stress months under both specifications, indicating that higher ESG is associated with lower crash-event incidence precisely when market conditions deteriorate. In the parsimonious model, the stress-month ESG coefficient is $-0.0844$ ($z=-3.44$, $p=0.001$), implying an odds ratio of $0.919$ per one-unit increase in ESG and $0.656$ per five-unit increase. In the specification with lagged firm controls, the coefficient remains negative at $-0.0765$ ($z=-3.01$, $p=0.003$), corresponding to odds ratios of $0.926$ per one-unit increase and $0.682$ per five-unit increase.

By contrast, the non-stress estimates are not comparably protective. The ESG slope is statistically insignificant in the parsimonious specification and turns positive once firm controls are added. The main implication is therefore not that ESG uniformly lowers crash risk in all months, but that its association with crash-event incidence is concentrated in adverse market states. Section~\ref{sec:robustness:threshold} shows that the same stress-state pattern remains under alternative crash thresholds.

\begin{table}[htbp]
\centering
\caption{Regime-Specific ESG Association in Crash-Event Incidence ($\crashprimary$).}
\label{tab:logit_crash20_interaction}
\scriptsize
\begin{tabular}{lcccc}
\toprule
& \multicolumn{2}{c}{Stress months} & \multicolumn{2}{c}{Non-stress months} \\
\cmidrule(lr){2-3}\cmidrule(lr){4-5}
& Spec.~A & Spec.~B & Spec.~A & Spec.~B \\
\midrule
ESG coefficient & $-0.0844^{***}$ & $-0.0765^{***}$ & 0.0498 & $0.0772^{**}$ \\
$z$-stat        & ($-3.44$)       & ($-3.01$)       & (1.38) & (2.07) \\
OR per +1 ESG   & 0.919           & 0.926           & 1.051  & 1.080 \\
OR per +5 ESG   & 0.656           & 0.682           & 1.283  & 1.471 \\
\bottomrule
\end{tabular}
\begin{minipage}{0.98\linewidth}
\vspace{0.15cm}
\footnotesize
\textit{Notes:} Entries are regime-specific ESG logit coefficients with month-clustered $z$-statistics in parentheses. Odds ratios are defined as $\exp(\hat\beta)$ for a one-unit increase in ESG and $\exp(5\hat\beta)$ for a five-unit increase. $^{**} p<0.05$, $^{***} p<0.01$.
\end{minipage}
\end{table}

\subsection{Excess-Return Tail Severity in Stress Regimes}
\label{sec:results:4b_quantile}

To locate where the lower crash-event incidence comes from, Table~\ref{tab:quantile_excess_tail} turns to the continuous excess-return distribution. It reports conditional quantile slopes for firm-month excess returns over a grid of left-tail quantiles.

First, the stress indicator corresponds to a materially worse downside environment even after subtracting the contemporaneous market return. At the 1st percentile, stress months shift the conditional left tail downward by $-0.064$ (95\% CI $[-0.177,-0.007]$), and a similar shift appears at the 2nd percentile ($-0.055$; 95\% CI $[-0.126,-0.006]$). The stress effect remains negative deeper in the distribution and is still statistically supported at $\tau=0.20$, where the coefficient is $-0.015$ (95\% CI $[-0.027,-0.003]$). This confirms that the stress classification captures a distinct cross-sectional downside regime rather than simply reflecting marketwide moves.

Second, within that downside regime, ESG protection is concentrated in the ultra-left tail rather than spread uniformly across the return distribution. The implied stress-month ESG slope $\beta_{\tau}+\delta_{\tau}$ is positive and statistically supported at crash-like quantiles: at $\tau=0.01$, the stress-month ESG effect is $0.006$ (95\% CI $[0.001,0.019]$), and at $\tau=0.02$ it is $0.006$ (95\% CI $[0.001,0.014]$). This protective association attenuates and becomes statistically weaker at higher quantiles (e.g., $\tau\in\{0.05,0.10,0.20\}$), while the non-stress ESG slope remains economically small throughout. The quantile evidence therefore sharpens Section~\ref{sec:results:crashlogit}: the stress-state crash protection is not a broad return effect, but is concentrated in the most adverse realized tail outcomes.

\begin{table*}[!h]
\centering
\caption{Conditional Quantiles of Monthly Excess Returns.}
\label{tab:quantile_excess_tail}
\smaller
\begin{tabular}{lcccc}
\toprule
Quantile $\tau$ &
Stress &
ESG (non-stress) &
ESG\,$\times$\,Stress &
ESG effect in stress \\
\midrule
0.01 &
$-0.0642^{*}$ [$-0.1773,\,-0.0068$] &
$-0.0000$ [$-0.0021,\,0.0025$] &
$0.0061$ [$-0.0002,\,0.0188$] &
$0.0061^{*}$ [$0.0011,\,0.0185$] \\
0.02 &
$-0.0548^{*}$ [$-0.1261,\,-0.0063$] &
$0.0004$ [$-0.0008,\,0.0022$] &
$0.0055^{*}$ [$0.0004,\,0.0132$] &
$0.0059^{*}$ [$0.0012,\,0.0138$] \\
0.05 &
$-0.0325$ [$-0.0723,\,0.0008$] &
$-0.0000$ [$-0.0009,\,0.0011$] &
$0.0028$ [$-0.0007,\,0.0072$] &
$0.0027$ [$-0.0006,\,0.0070$] \\
0.10 &
$-0.0151$ [$-0.0442,\,0.0001$] &
$-0.0001$ [$-0.0008,\,0.0006$] &
$0.0006$ [$-0.0009,\,0.0042$] &
$0.0004$ [$-0.0008,\,0.0040$] \\
0.20 &
$-0.0147^{*}$ [$-0.0266,\,-0.0026$] &
$-0.0004$ [$-0.0008,\,0.0002$] &
$0.0010$ [$-0.0004,\,0.0025$] &
$0.0007$ [$-0.0006,\,0.0019$] \\
\bottomrule
\end{tabular}

\begin{minipage}{\textwidth}
\vspace{0.1cm}
\footnotesize
\textit{Notes:} The outcome is monthly excess return. All specifications include sector fixed effects and lagged firm controls. Brackets report 95\% confidence intervals from a stratified month-block bootstrap (stress and non-stress months resampled separately; 800 replicates). $^{*}$ indicates the 95\% interval excludes zero.
\end{minipage}
\end{table*}

\subsection{Deconfounded ESG Effects Across Stress States}
\label{subsec:dml-result}

In this section, we examine the after treatment effect from the DML approach, as is introduced in \Cref{subsec:dml-setup}.
Table \ref{tab:result-55-dml_summary} confirms that the ESG payoff ($\beta$) is structurally asymmetric and highly state-dependent.
Utilizing Double Machine Learning (DML) to control for firm characteristics, we find that ESG scores offer no predictive value for crash risk or excess returns during non-stress regimes, with estimates statistically indistinguishable from zero.

Conversely, a robust interaction effect emerges during stress regimes:
\begin{itemize}
    \item \textbf{Downside Risk}: ESG significantly mitigates crash probability, supported by both Lasso ($z = -5.53, p < 0.01$) and Random Forest ($z = -4.44, p < 0.01$) models.
    \item \textbf{Excess Returns}: Results are inconsistent; while the Lasso estimator shows a marginal positive effect ($z = 1.96, p < 0.05$), the Random Forest estimate ($z = 1.42$) fails to reach significance.
\end{itemize}
These findings underscore that high ESG ratings function primarily as a downside protection mechanism, which acts as priced "crash insurance" rather than a reliable driver of alpha during market turbulence.

\begin{table*}[htbp]
\centering
\caption{Double Machine Learning Estimates of ESG Score Treatment Effect: Stress vs. Non-stress Regimes.}
\label{tab:result-55-dml_summary}
\small
\begin{tabular}{@{}l cccc @{}}
\toprule
 & \multicolumn{2}{c}{\textbf{Stress Regime}} & \multicolumn{2}{c}{\textbf{Non-stress Regime}} \\
\cmidrule(lr){2-3} \cmidrule(l){4-5}
\textbf{Outcome Variable} & \textbf{Lasso} & \textbf{Random Forest} & \textbf{Lasso} & \textbf{Random Forest} \\
\midrule

Panel A: Downside Risk ($\crashprimary = 1$) & $-0.0039^{***}$\ $(-5.53)$ & $-0.0032^{***} $\  $(-4.44)$ & $0.0000$ \  $(0.12)$ & $0.0000$ \ $(0.42)$ \\
\addlinespace

Panel B: Excess Return ($\retex$) & $0.0006^{**}$ \ $(1.96)$ & $0.0004 $\  $(1.42)$ & $-0.0002$ \  $(-1.52)$ & $-0.0002$ \ $(-1.34)$ \\

\bottomrule
\end{tabular}
\begin{minipage}{\textwidth}
\vspace{0.1cm}
\footnotesize
\textit{Notes:} This table presents the average treatment effect of the Industry Adjusted ESG Score on crash indicator and excess returns using Double Machine Learning (DML). The sample is partitioned into stress and non-stress regimes. $z$-statistics are reported in parentheses. Significance levels: $^{*} p < 0.10$, $^{**} p < 0.05$, $^{***} p < 0.01$.
\end{minipage}
\end{table*}

\section{Robustness Checks}
\subsection{Crash Threshold Sensitivity}
\label{sec:robustness:threshold}

Table~\ref{tab:threshold_robustness} shows that the main conclusion (Section~\ref{sec:results:crashlogit}) is not driven by the headline crash cutoff. We repeat the analysis for $\crash^{(0.15)}_{i,t}$, $\crash^{(0.20)}_{i,t}$, and $\crash^{(0.25)}_{i,t}$, holding the stress classification fixed throughout.

The descriptive pattern is stable in stress months: crash incidence declines from low-ESG to high-ESG firms at all three thresholds, and the Q1--Q5 spread remains statistically supported under month-block bootstrap inference. The stress-month Q1--Q5 gap is 5.763 for $\crash^{(0.15)}$, 3.532 for $\crash^{(0.20)}$, and 2.354 for $\crash^{(0.25)}$.

The multivariate evidence points in the same direction. In the preferred specification with controls, the ESG slope is negative in stress months for all three cutoffs: $-0.0457$ ($p=0.029$) for $\crash^{(0.15)}$, $-0.0765$ ($p=0.003$) for $\crash^{(0.20)}$, and $-0.0643$ ($p=0.083$) for $\crash^{(0.25)}$. The implied odds ratio per five-unit increase in ESG is 0.796, 0.682, and 0.725, respectively. Outside stress months, the ESG slope is not comparably protective. Overall, the regime-specific crash-logit evidence remains concentrated in stress months, with the headline $\crash^{(0.20)}$ cutoff delivering the sharpest multivariate result.



\begin{table}[htbp]
\centering
\caption{Robustness to Alternative Crash Thresholds.}
\label{tab:threshold_robustness}
\scriptsize

\begin{minipage}[t]{0.49\textwidth}
\centering
\textbf{Panel A: Descriptive Stress-Month ESG Gradient}

\vspace{0.6em}
\begin{tabular}{lccc}
\toprule
 & $\crash^{(0.15)}$ & $\crash^{(0.20)}$ & $\crash^{(0.25)}$ \\
\midrule
Stress-month crash rate & 10.520\% & 4.483\% & 2.242\% \\
Non-stress crash rate   & 1.521\%  & 0.602\% & 0.285\% \\
Q1 crash rate in stress & 12.895\% & 6.105\% & 3.158\% \\
Q5 crash rate in stress & 7.131\%  & 2.574\% & 0.804\% \\
Q1--Q5 gap              & 5.763    & 3.532   & 2.354 \\
95\% CI for gap         & [1.640, 11.122] & [1.008, 7.533] & [0.485, 5.515] \\
\bottomrule
\end{tabular}
\end{minipage}
\hfill

\vspace{0.8em}

\begin{minipage}[t]{0.49\textwidth}
\centering
\textbf{Panel B: Regime-Specific Logit Robustness (Spec.~B)}

\vspace{0.6em}
\begin{tabular}{ccccc}
\toprule
Regime & Statistic & $\crash^{(0.15)}$ & $\crash^{(0.20)}$ & $\crash^{(0.25)}$ \\
\midrule
\multirow{3}{*}{Stress}
& ESG coef.     & $-0.0457^{**}$ & $-0.0765^{***}$ & $-0.0643^{*}$ \\
& $p$-value     & 0.029          & 0.003           & 0.083 \\
& OR per +5 ESG & 0.796          & 0.682           & 0.725 \\
\cmidrule(lr){1-5}
\multirow{3}{*}{Non-stress}
& ESG coef.     & 0.0068         & $0.0772^{**}$   & 0.0527 \\
& $p$-value     & 0.780          & 0.038           & 0.297 \\
& OR per +5 ESG & 1.035          & 1.471           & 1.301 \\
\bottomrule
\end{tabular}
\end{minipage}

\begin{minipage}{0.49\textwidth}
\vspace{0.4em}
\footnotesize
\textit{Notes:} Panel A reports descriptive crash rates and the Q1--Q5 ESG spread within stress months, with 95\% month-block bootstrap confidence intervals. Panel B reports regime-specific ESG slopes from the preferred crash logit with lagged firm controls and sector fixed effects. Standard errors are clustered by month. $^{*} p<0.10$, $^{**} p<0.05$, $^{***} p<0.01$.
\end{minipage}
\end{table}

\subsection{Pillar Score Effects}
\label{subsec:pillar-dml}
Following results in \cref{subsec:dml-result}, we wish to further examine the effectiveness of pillar scores using the DML approach.
Decomposing the treatment effect reveals that while all pillars perform relatively better during crises, their underlying mechanics differ, as is contrasted in \Cref{fig:result-55-dml-coef-forest-both}:

\begin{figure}[htbp]
    \centering

    \begin{subfigure}[t]{0.95\linewidth}
        \centering
        \includegraphics[width=\linewidth]{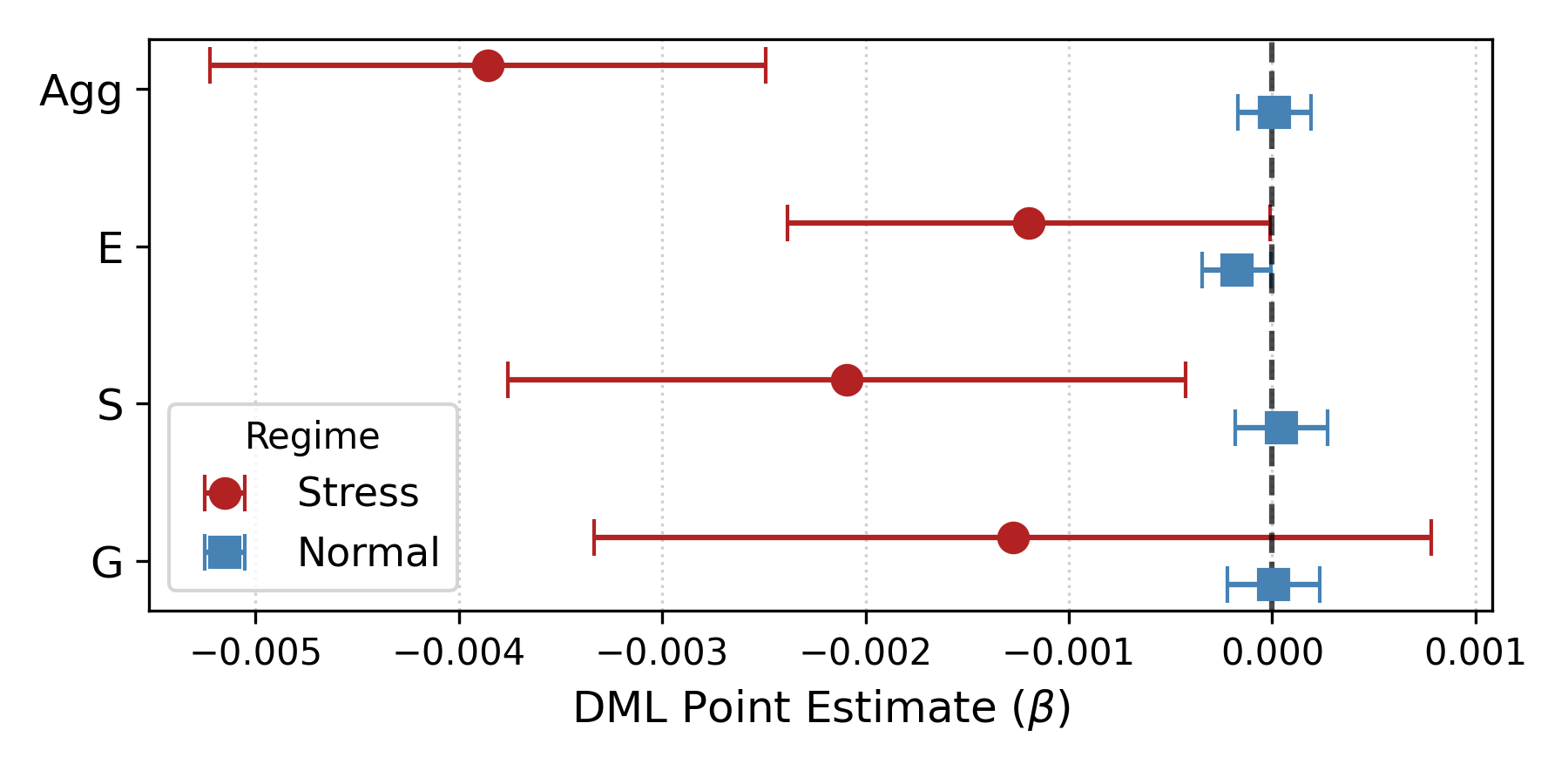}
        \caption{Target: $\crashprimary$}
        \label{fig:result-55-dml-coef-forest-crash}
    \end{subfigure}\hfill
    \begin{subfigure}[t]{0.95\linewidth}
        \centering
        \includegraphics[width=\linewidth]{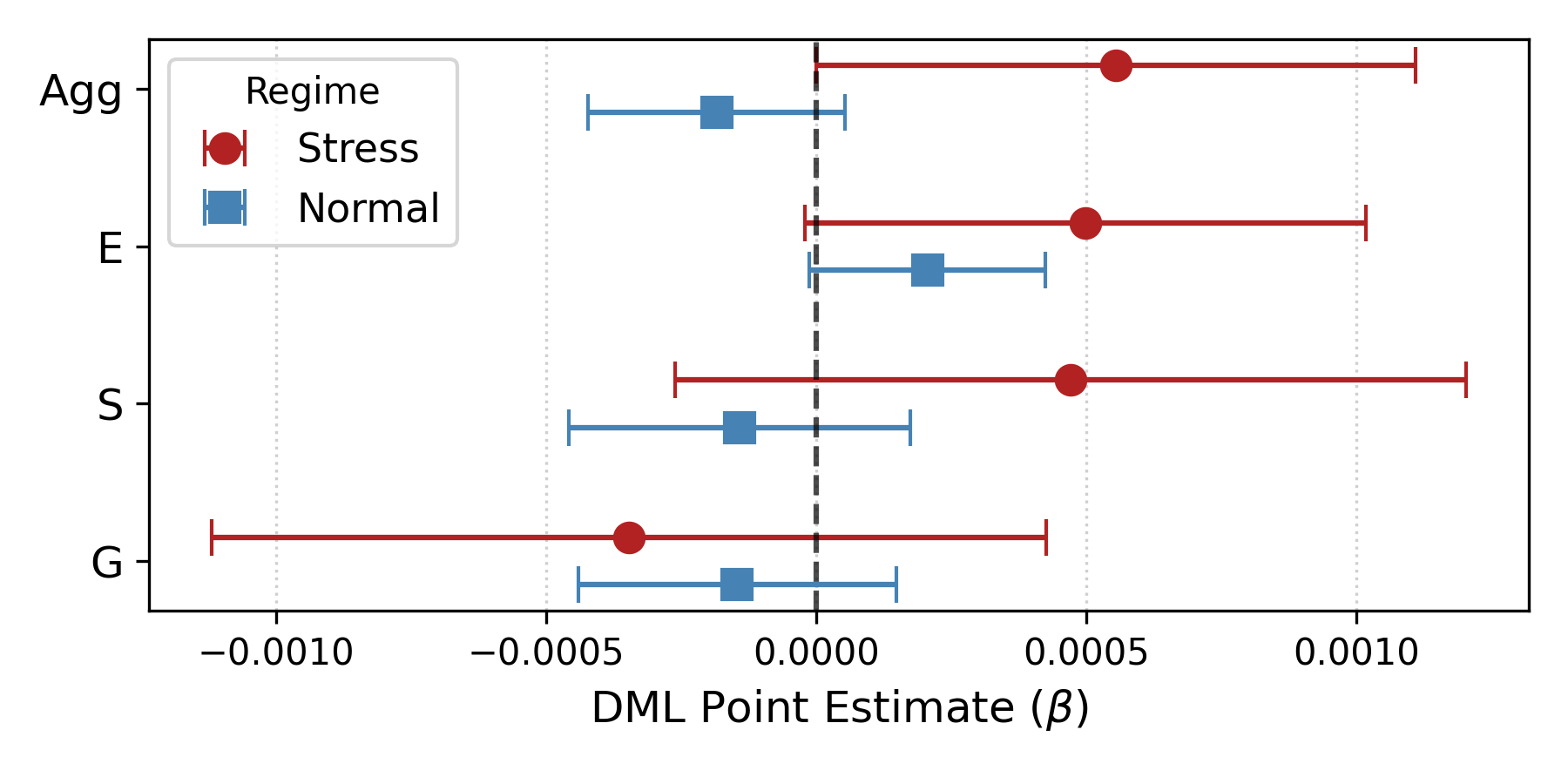}
        \caption{Target: $\retex$}
        \label{fig:result-55-dml-coef-forest-ret}
    \end{subfigure}

    \caption{DML Treatment Effects of ESG Components by Market Regimes.}
    \label{fig:result-55-dml-coef-forest-both}
    
\vspace{10pt}
\raggedright
\small
\textit{Notes:} This figure illustrates the after-treatment effect estimated via the Double Machine Learning (DML) approach under a Lasso specification. 
The \textit{Agg} variable represents the Industry Adjusted Aggregated ESG Score. 
Components are defined as: 
\textbf{E}: Environmental Score; 
\textbf{S}: Social Score; 
\textbf{G}: Governance Score. 
Point estimates ($\beta$) and 95\% confidence intervals are partitioned by market regime, where \textit{Stress} (red) denotes periods of high volatility and \textit{Normal} (blue) denotes non-stress periods.
\end{figure}

\begin{enumerate}
    \item[\textbf{S}] The Social pillar is the most consistent driver of performance during crises, functioning as a critical proxy for implicit incorporeal assets such as stakeholder trust. It significantly reduces crash probability during stress regimes while simultaneously serving as the primary contributor to positive excess returns during those same periods. This suggests that factors like workforce stability and supply chain trust provide a unique form of tangible resilience when macroeconomic volatility spikes.
    \item[\textbf{E}] The Environmental pillar best illustrates the "cost" of ESG insurance. In normal regimes, high E-scores are associated with a slight drag on returns. However, during stress regimes, this pillar provides substantial downside protection by significantly reducing the likelihood of extreme drawdowns. This confirms that green investments act as a defensive hedge that pays off in relative terms during systemic market crashes.
    \item[\textbf{G}] Finally, the Governance pillar exhibits the weakest marginal effect across both regimes in our DML specification. This muted G-pillar effect is economically intuitive because our DML first-stage rigorously partials out fundamental accounting characteristics (e.g., profitability, leverage, size), much of the variance typically associated with "good governance" has already been absorbed by our high-dimensional controls.
\end{enumerate}

\section{Conclusion}

This paper characterizes ESG as a state-dependent insurance mechanism rather than a source of persistent alpha. Our drawdown-based stress rule provides a validated state object for analysis by identifying abrupt nonlinear market transitions. We show that high ESG ratings reduce discrete crash incidence during turbulent periods while offering no comparable protection in stable markets. Using Double Machine Learning as a structural deconfounding layer, we address selection bias and high-dimensional confounding to isolate the asymmetric ESG effect across states. Tail-severity analysis further shows that this protection is concentrated in the extreme lower tail during stress months. Overall, the evidence indicates that ESG is priced less as a source of unconditional return enhancement than as a form of crash insurance whose value is realized when systemic fragility is highest. Future research may extend this framework to global ESG ratings and to the macroeconomic implications of ESG-augmented risk management for market equilibrium.

\begin{acks}
During the preparation of this work the author(s) used ChatGPT (GPT-5, OpenAI) and Gemini in order to perform a language check. The authors reviewed and edited the content as needed and take full responsibility for the content.
\end{acks}

\bibliographystyle{ACM-Reference-Format}
\bibliography{ref}

\appendix

\end{document}